\documentclass[aps,prb,twocolumn,groupedaddress]{revtex4-2}

\usepackage[T1]{fontenc}
\usepackage{amsmath}
\usepackage{amssymb}
\usepackage{bbm}
\usepackage{xspace}
\usepackage{color}
\usepackage{xcolor}
\usepackage{graphicx}
\usepackage{ulem} 
\usepackage{placeins}
\usepackage[colorlinks=true,linkcolor=blue,citecolor=blue,urlcolor=blue]{hyperref}
\usepackage{bm}
\usepackage{lipsum}

\DeclareMathOperator{\arccosh}{arcosh}

\graphicspath{{.}{Bilder/}}

\begin{document}

\title{Light-modulated Josephson effect in Kekul\'e patterned graphene}
\author{W.~Zeng$^1$}
\author{R.~Shen$^{1,2}$}
\email[E-mail: ]{shen@nju.edu.cn}
\affiliation{$^1$National Laboratory of Solid State Microstructures and School of Physics, Nanjing University, Nanjing 210093, China\\
$^2$Collaborative Innovation Center of Advanced Microstructures, Nanjing University, Nanjing, 210093, China}

\date{\today}

\begin{abstract}{}
We theoretically study the Josephson effect in a superconductor/normal metal/superconductor junction based on Kekul\'e  patterned graphene. For the Kekul\'e-O patterned junctions, a Fermi momentum-splitting Andreev reflection at the interface can be induced by the off-resonant circularly polarized light applied in the normal region, which results in the possible $\pi$-state. In contrast, for the Kekul\'e-Y patterned junctions, the Fermi momentum-splitting Andreev reflection is strongly suppressed due to the valley-momentum locking and the junction always exhibits the 0-state. The dependence of the critical current on the junction length and the illumination parameter of the light field is also presented in detail.
\end{abstract}

\maketitle
\section{Introduction}\label{intro}

Kekul\'e (Kek) patterned graphene is a two-dimensional superlattice consisting of periodic thin and thick atomic bonds\cite{PhysRevB.80.233409,PhysRevB.62.2806,PhysRevLett.98.186809,CHEIANOV20091499}. These alternating bonds form a $\sqrt{3}\times\sqrt{3}$ supercell, where the $K$ and $K'$ valleys of the pristine graphene are folded on top of each other\cite{Gamayun_2018,PhysRevB.100.075431,PhysRevLett.126.206804}. Recent experiments demonstrated that a carbon atom-centered bond texture can be realized in the graphene sheet grown on Cu(111)\cite{gutierrez2016imaging,CHEIANOV20091499}, which is known as the Kek-Y patterned graphene. The valley-momentum locking\cite{Gamayun_2018} predicted in this system leads to many peculiar properties, such as the enhanced Andreev reflection\cite{PhysRevB.104.075436}, the valley precession effect\cite{PhysRevB.98.195436}, the resonant transport\cite{andrade2020resonant}, and the tunable optical absorption\cite{PhysRevB.101.205413}. However, the pattern in which the C-C bond strength is altered as in a benzene ring is known as the Kek-O bond texture with a gap opened at the Dirac point\cite{Gamayun_2018}. Recently, in the Kek-O patterned graphene, Beenakker \textit{et al.} predicted a valley switch effect by use of the Andreev-like reﬂection\cite{PhysRevB.97.241403} and Wang \textit{et al.} subsequently reported the valley supercurrent\cite{PhysRevB.101.245428}. 

The Josephson effect is an example of a macroscopic quantum phenomenon first predicted by B. D Josephson in 1962\cite{josephson1962possible,RevModPhys.46.251}. A Josephson junction consists of two superconductors coupled by a weak link\cite{PhysRevB.74.041401,PhysRevB.95.064511,PhysRevB.102.085144,PhysRevB.103.125147,PhysRevB.98.075430}. The current flowing continuously across the junction without any voltage applied is called the dc Josephson current\cite{FURUSAKI1991299,PhysRev.175.585,PhysRevB.2.2543,PhysRevLett.21.1241}, which is driven by the superconducting phase difference $\phi$ and can be expressed as $I\sim\sin(\phi+\phi_0)$ with $\phi_0$ representing the additional phase shift. The ground state of the junction usually has a zero phase shift at $\phi_0=0$ due to the time-reversal symmetry\cite{RevModPhys.76.411,RevModPhys.77.935,RevModPhys.77.1321}. With a ferromagnetic link, an extra $\pi$ phase shift can appear in the junction leading to the $\pi$-state junctions with the supercurrent reversals\cite{PhysRevLett.86.2427,PhysRevLett.89.137007,zeng20180,PhysRevLett.66.3056,PhysRevLett.67.3836}. As an analogy to the spin polarization, several studies have revealed that the valley isospin polarizations in graphene-like materials can also result in the $\pi$-state junctions, such as the irradiated graphene- and silicene-based Josephson junctions\cite{PhysRevB.94.165436,PhysRevB.89.064501}.

However, the discussions on the Andreev reflection and the Josephson effect in Kek patterned graphene are still insufficient in the literature. Very recently, Mojarro \textit{et al.} studied the dc conductivity of the Kek patterned graphene under the circularly polarized light\cite{PhysRevB.102.165301}. Motivated by this, we report a study on the light-modulated Josephson effect in Kek patterned graphene in this paper. For the Kek-O patterned Josephson junctions, it is found that the valley-degenerate band is split into two parts with opposite valley polarization in the presence of a light field, leading to the $\pi$-state junctions. In contrast, the $\pi$-state is always absent in Kek-Y patterned junctions due to the valley-momentum locking. The critical current dependence on the junction length and the illumination parameter of the light field is also presented.

The rest of this paper is organized as follows. The model Hamiltonian and the scattering approach are explained in detail in Sec.\ \ref{model}. The numerical results and discussions are presented in Sec.\ \ref{results}. Finally, we conclude in Sec.\ \ref{conclusions}.

\section{theoretical Model}\label{model}
\subsection{Hamiltonian}

\begin{figure}[tb]
\centerline{\includegraphics[width=1\linewidth]{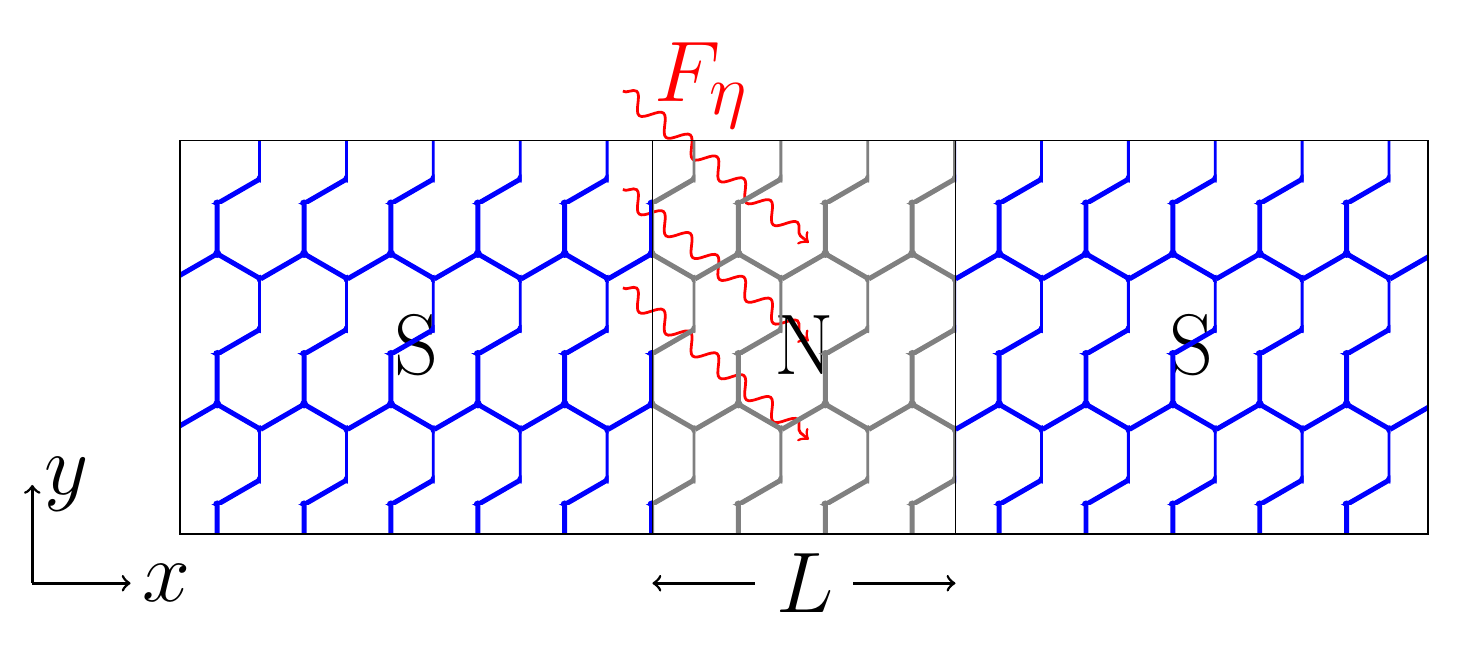}}
\caption{\label{f0}
Schematic diagram of the junction. The blue and the gray regions indicate the superconducting and the normal regions, respectively. The off-resonant polarized light applied in the normal region is indicated by the red wavy lines.
}
\end{figure}

We consider the superconductor/normal metal/superconductor junction based on the Kek patterned graphene, as shown in Fig.\ \ref{f0}. The normal region at $0<x<L$ is irradiated by an off-resonant circularly polarized light. The two superconducting regions are realized by the superconducting proximity effect\cite{VOLKOV1995261,efetov2016specular}.
The low energy Hamiltonian of the Kek graphene reads\cite{Gamayun_2018} 
\begin{align}
 &\mathcal{H}_0=\hbar v_F\tau_0\bm k\cdot\bm\sigma+\mathcal{H}'-U,\label{h0}\\
&\mathcal{H}'=\begin{cases}
 \mathcal{C}_O\tau_x\sigma_z& \text{Kek-O}\\ 
 \mathcal{C}_Y\hbar v_F\tilde{\bm k}\cdot\bm \tau\sigma_0& \text{Kek-Y}
\end{cases},\label{hkek}
\end{align} 
where $(\psi_{AK},\psi_{BK},-\psi_{BK'},\psi_{AK'})^T$ is the valley isotropic basis\cite{Gamayun_2018,RevModPhys.80.1337} with $A,B$ and $K,K'$ indicating the sublattices and the valleys, respectively, $v_F$ is the Fermi velocity, $\bm \sigma$ and $\bm \tau$ are the Pauli matrices acting on the sublattice and the valley space, respectively. The momentum is defined as $\bm k(\tilde{\bm k})=(\hat{k}_x,\pm \hat{k}_y)$ with $\hat{k}_{x(y)}=-i\partial_{x(y)}$. The electrostatic potential $U$ is zero in the normal region and is finite in the superconducting region, which can be adjusted by doping or by a gate voltage.  The coupling amplitude $\mathcal{C}_{O(Y)}$ is introduced by the Kek-O (-Y) bond density wave, which can be adopted as a real number under an appropriate unitary transformation\cite{Gamayun_2018}. We note that the Kek-Y Hamiltonian $\mathcal{H}'$ in Eq.\ (\ref{hkek}) is valid in small-$\mathcal{C}_Y$ regime ($\mathcal{C}_Y\ll1$), otherwise the renormalization effect of the Fermi velocities should be considered\cite{Gamayun_2018}.

When a beam of off-resonant circularly polarized light is uniformly launched into the normal region, the electrons could feel a time-dependent vector potential $\bm A=A(\eta\sin\omega t,\cos\omega t)$, where $\eta=\pm 1$ denotes the right (left) circularly polarized light, $A$ and $\omega$ are the amplitude and the frequency of the light, respectively. By substituting $\bm k$ with $\bm k+e\bm A/\hbar$, the Floquet Hamiltonian in the regime of $\hbar\omega\gg eAv_F$ is obtained as\cite{PhysRevB.84.235108,cayssol2013floquet} 
\begin{align}
\mathcal{H}_{\mathrm{F}}(\boldsymbol{k}) \simeq \mathcal{H}_{0}(\boldsymbol{k})+\frac{\left[\mathcal{H}_{-1}, \mathcal{H}_{+1}\right]}{\hbar \omega},\label{ef1}
\end{align}
with
\begin{align}
\mathcal{H}_{m}(\boldsymbol{k})=\frac{1}{T} \int_{0}^{T} \mathrm{d} t e^{i m \omega t} \mathcal{H}(\boldsymbol{k}, t),\ (m=0,\pm1)
\end{align}
and $T=2\pi/\omega$ being the time period. By substituting Eq.\ (\ref{h0}) into Eq.\ (\ref{ef1}), one obtains
\begin{align}
\mathcal{H}_F=\begin{cases}
F_\eta\tau_0\sigma_z & \text{Kek-O } \\ 
F_\eta\tau_0\sigma_z-\mathcal{C}_Y^2F_\eta\tau_z\sigma_0 & \text{Kek-Y }
\end{cases}\label{flo},
\end{align}
where $F_\eta=8\pi\alpha\eta P/\omega^3$ is the illumination parameter and $P=(eA\omega)^2/8\pi\alpha$ is the laser intensity with $\alpha\simeq 1/137$ being the ﬁne-structure constant, respectively. The total Hamiltonian in the normal region is given by 
\begin{align}
\mathcal{H}=\mathcal{H}_0+\mathcal{H}_F.\label{total}
\end{align}

The electron transport properties are studied under the Dirac-Bogoliubov–de Gennes (DBdG) approach\cite{de2018superconductivity}. The DBdG equation reads
\begin{align}
\begin{pmatrix}
\mathcal{H}-\mu &\Delta(x) \\ 
\Delta^*(x) &\mu-\mathcal{T}\mathcal{H}\mathcal{T}^{-1} 
\end{pmatrix}\begin{pmatrix}
\psi_e\\ 
\psi_h
\end{pmatrix}=\varepsilon\begin{pmatrix}
\psi_e\\ 
\psi_h
\end{pmatrix},\label{bdg}
\end{align}
where $\varepsilon$ is the excitation energy measured from the Fermi level $\mu$, $\psi_{e}$ ($\psi_h$) is the electron- (hole-) component of the quasiparticle wave function, and $\mathcal{T}=-\tau_y\sigma_y\mathcal{K}$ is the time-reversal operator with $\mathcal{K}$ being the complex conjugation operator, respectively. Since the superconductor region is also Kek graphene based, the lattice mismatch is minimal and the interface barrier is omitted in our model. Under the widely adopted heavy doping condition, the Fermi wave vector in the superconductor is much larger than that in the normal metal so that the s-wave pair potential $\Delta(x)$ can be approximated as a steplike function\cite{PhysRevB.82.115437,RevModPhys.80.1337,PhysRevLett.97.067007}, i,e., $\Delta(x)=\Delta e^{i\phi/2}$ for $x<0$ and $\Delta(x)=\Delta e^{-i\phi/2}$ for $x>L$ with $\phi$ being the phase difference across the junction. The temperature-dependence of the gap function is $\Delta=\Delta_0\tanh(1.74\sqrt{T_c/T-1})$\cite{tinkham2004introduction}, where $\Delta_0$ is the gap at the zero temperature and $T_c$ is the superconducting transition temperature, respectively. Since the two spin channels are decoupled, there is no spin indices in DBdG Eq.\ (\ref{bdg}) and the single particle Hamiltonian takes a $4\times4$ form with the indices of the sublattice pseudospin and the valley isospin.

\subsection{Scattering approach}
\subsubsection{Andreev reflection matrix $\mathcal{U}$}
Following the approach in Refs.\ \cite{PhysRevB.74.041401,PhysRevB.95.064511}, we derive the Andreev reflection matrix at $x=0$. Under the heavily doping condition, the eigenstates of Eq.\ (\ref{bdg}) in the region $x<0$ can be obtained as
\begin{widetext}
\begin{align}
\psi_{a\xi}^O(x)=&\begin{pmatrix}
e^{i\frac{\phi}{2}} &\xi e^{i\frac{\phi}{2}}  &0  &0&e^{i\xi\beta}&\xi e^{i\xi\beta}&0&0 
\end{pmatrix}^T\times e^{i\xi k^Ox+\kappa^Ox},\label{fs}\\
\psi_{b\xi}^O(x)=&\begin{pmatrix}
0&0&e^{i\frac{\phi}{2}} &\xi e^{i\frac{\phi}{2}}&0&0&e^{i\xi\beta}&\xi e^{i\xi\beta} 
\end{pmatrix}^T\times e^{i\xi k^Ox+\kappa^Ox}.\label{fs2}
\end{align}
for the Kek-O junctions and 
\begin{align}
\psi_{a\xi}^Y(x)=&\begin{pmatrix}
\xi e^{i\frac{\phi}{2}} &e^{i\frac{\phi}{2}}  &e^{i\frac{\phi}{2}}  &\xi e^{i\frac{\phi}{2}}&\xi e^{i\xi\beta}&e^{i\xi\beta}&e^{i\xi\beta}&\xi e^{i\xi\beta} 
\end{pmatrix}^T \times e^{i\xi k_a^Yx+\kappa_a^Yx},\label{fs3}\\
\psi_{b\xi}^Y(x)=&\begin{pmatrix}
-\xi e^{i\frac{\phi}{2}} &-e^{i\frac{\phi}{2}}  &e^{i\frac{\phi}{2}}  &\xi e^{i\frac{\phi}{2}}&-\xi e^{i\xi\beta}&-e^{i\xi\beta}&e^{i\xi\beta}&\xi e^{i\xi\beta} 
\end{pmatrix}^T \times e^{i\xi k_b^Yx+\kappa_b^Yx},\label{fs4}
\end{align}

\end{widetext}
for the Kek-Y junctions, where $\xi=\pm 1$, $k^O=U/\hbar v_F$, $k_{a(b)}^{Y}=U/[\hbar v_F(1\mp\mathcal{C}_Y)]$, $\kappa^O=\Delta_0\sin\beta/\hbar v_F$, $\kappa_{a(b)}^Y=\Delta_0\sin\beta/[\hbar v_F(1\mp\mathcal{C}_Y)]$, $\beta=\arccos(\varepsilon/\Delta_0)$ for $\varepsilon<\Delta_0$ and $\beta=-i\arccosh(\varepsilon/\Delta_0)$ for $\varepsilon>\Delta_0$, respectively.

At $x=0$, the electron component $\psi_e(0)$ and the hole component $\psi_h(0)$ of the superconducting wave function are connected by the Andreev reflection matrix $\mathcal{U}$ as 
\begin{align}
\psi_h(0)=\mathcal{U}\psi_e(0),\label{rl}
\end{align}
where $\psi_e(0)$ and $\psi_h(0)$ are the $4\times1$ vectors and $\mathcal{U}$ is a $4\times4$ matrix. With the help of the four wave functions in Eqs.\ (\ref{fs}) and (\ref{fs2}), the Andreev reflection matrix for the Kek-O junction can be obtained. Similarly, the Andreev reflection matrix for the Kek-Y junction can be obtained by substituting Eqs.\ (\ref{fs3}) and (\ref{fs4}) into Eq.\ (\ref{rl}). Finally, one obtains 
\begin{align}
\mathcal{U}_O&=\mathcal{U}_Y=\mathcal{U}\\\nonumber&=\frac{e^{-i\frac{\phi}{2}}}{\Delta_0}\mathcal{I}_{2\times2}\bigotimes\begin{pmatrix}
\varepsilon &i\sqrt{1-\varepsilon^2} \\ 
i\sqrt{1-\varepsilon^2} & \varepsilon
\end{pmatrix},
\end{align}
where $\mathcal{I}_{2\times2}$ is the $2\times2$ identity matrix acting on the valley space.

\subsubsection{Normal transfer matrices $\mathcal{M}_e$ and $\mathcal{M}_h$}

The transfer matrix in the normal region can be obtained by matching the normal state wave functions\cite{PhysRevB.97.241403,PhysRevB.96.035437,PhysRevLett.96.246802,PhysRevB.78.045118}. Taking the Kek-O junction as an example, the single particle wave function with the fixed transverse wave vector $q$ and the excitation energy $\varepsilon$ is expressed as $\Psi(x,y)=\Psi_q(x)e^{iqy}$, which satisfies the Schrodinger equation $\mathcal{H}\Psi(x,y)=\varepsilon\Psi(x,y)$. With the help of Eq.\ (\ref{total}), one obtains 
\begin{align}
\frac{\partial\Psi_q(x)}{\partial x}=\frac{i}{\hbar v_F}\left(\tau_0\sigma_x\right)^{-1}\left[\varepsilon-\mathcal{H}(0,q)\right]\Psi_q(x).\label{dp}
\end{align}
The electron wave function at $x=0$ and that at $x=L$ are connected by  
\begin{align}
\Psi_q(L)=\mathcal{M}_e\Psi_q(0).
\end{align}
By integrating Eq.\ (\ref{dp}), the electron transfer matrix $\mathcal{M}_e$ is obtained as 

\begin{align}
\mathcal{M}_e&=e^{\Xi_e(\varepsilon,q)L},\\
\Xi_e(\varepsilon,q)=\frac{i}{\hbar v_F}&\left(\frac{1}{\varepsilon-\mathcal{H}(0,q)}\tau_0\sigma_x\right)^{-1}.
\end{align}
Similarly, the hole transfer matrix is obtained as 

\begin{align}
\mathcal{M}_h&=e^{\Xi_h(\varepsilon,q)L},\\
\Xi_h(\varepsilon,q)=\frac{i}{\hbar v_F}&\left(\frac{1}{\varepsilon-\mathcal{T}\mathcal{H}(0,-q)\mathcal{T}^{-1}}\tau_0\sigma_x\right)^{-1}.
\end{align}

\subsubsection{Josephson current}

The Josephson current at ﬁnite temperature is given by\cite{brouwer1997anomalous}
\begin{align}
I=-k_{\mathrm{B}} T \frac{4 e}{\hbar} \frac{\mathrm{d}}{\mathrm{d} \phi} \int_{0}^{\infty} d \varepsilon \varrho(\varepsilon) \ln \left[2 \cosh \left(\frac{\varepsilon}{2 k_{\mathrm{B}} T}\right)\right],\label{IF}
\end{align}
where $\varrho(\varepsilon)=\sum_i\delta(\varepsilon-\varepsilon_i)$ is the density of states of the Andreev levels. The discrete Andreev level $\varepsilon_i$ is determined by the secular equation\cite{PhysRevB.74.041401}
\begin{align}
\det\left(1-\mathcal{M}_e^{-1}\mathcal{U}\mathcal{M}_h\mathcal{U}\right)=0,\label{FM}
\end{align}
which comes from the fact that the Andreev process between the two normal metal/superconductor boundaries forms a closed loop. By an analytic continuation, the integration in Eq.\ (\ref{IF}) can be transformed into a summation over the Matsubara frequencies and one obtains 
\begin{align}
I=-\frac{4 e}{\hbar}  k_{\mathrm{B}} T \sum_{n,q} \frac{\mathrm{d}}{\mathrm{d} \phi} \ln \det\left(1-\mathcal{M}_e^{-1}\mathcal{U}\mathcal{M}_h\mathcal{U}\right),\label{josephson}
\end{align}
where the energy variable in the determinant is replaced by $i\omega_n$ with $\omega_n=(2n+1)\pi k_{\mathrm{B}}T$.

\section{Results}\label{results}
In the following calculations, we set $\Delta_0=1$ as the energy unit. The superconducting coherence length is $\xi_0=\hbar v_F/\Delta_0$. The Josephson current $I$ is renormalized by the current in the normal state $I_0=e\mu\Delta_0 W/\hbar^2\pi v_F$ with $W$ being the junction width. 

\begin{figure}[tb]
\centerline{\includegraphics[width=1\linewidth]{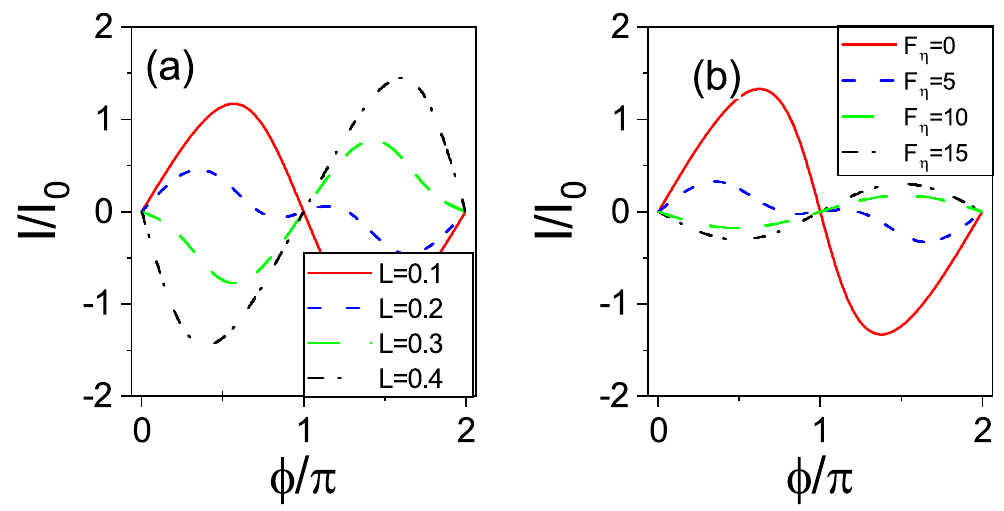}}
\caption{\label{f1}
Josephson current in Kek-O junctions with the parameters $\mu=42$, $\mathcal{C}_O=30$, $T/T_c=0.1$. (a) Current phase relation with $F_\eta=10$. (b) Current phase relation with $L=0.1$.
}
\end{figure}

\begin{figure}[tb]
\centerline{\includegraphics[width=1\linewidth]{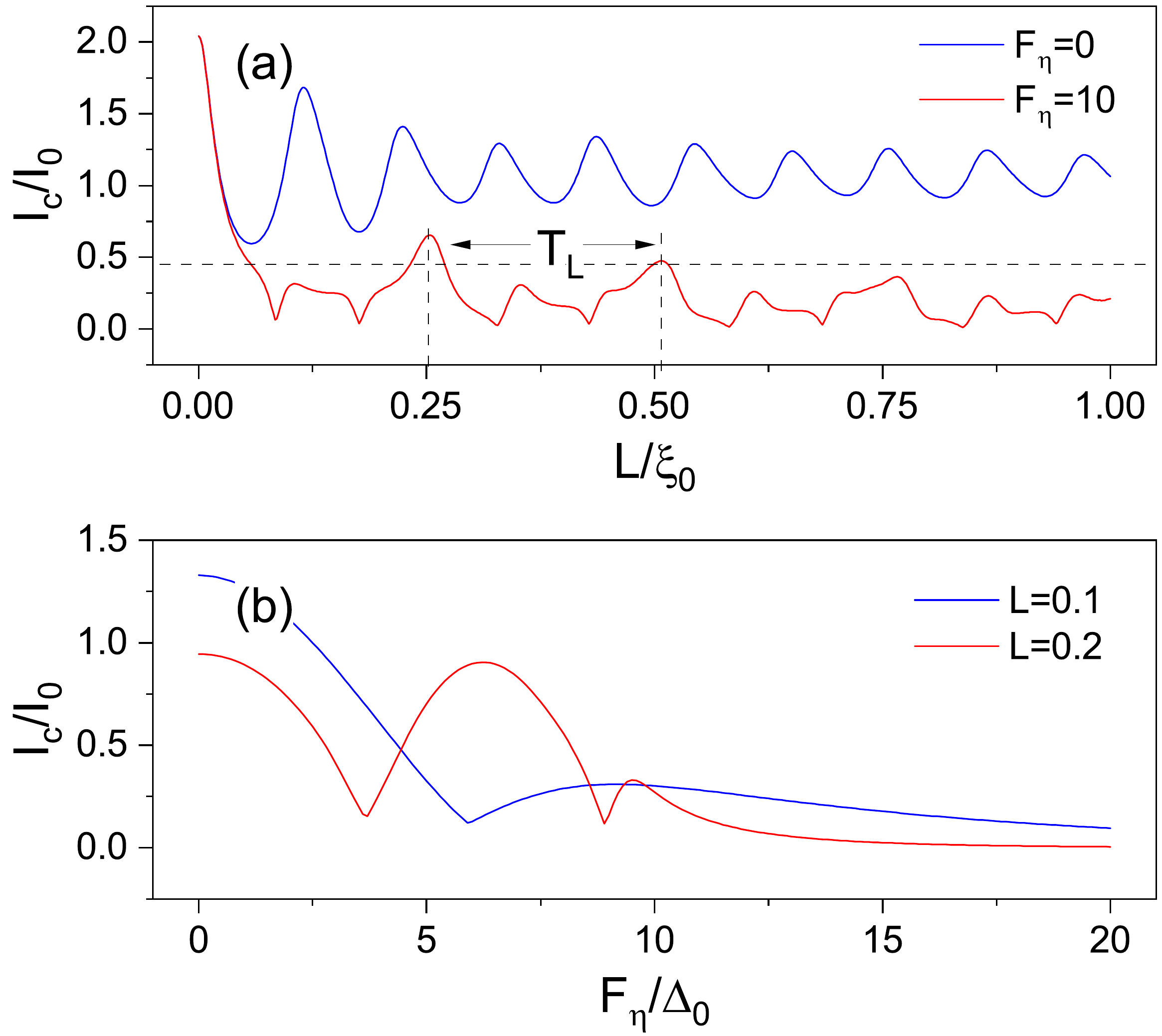}}
\caption{\label{f22}
Critical current in Kek-O junctions with the parameters $\mu=42$, $\mathcal{C}_O=30$, $T/T_c=0.1$. (a) Critical current as a function of $L$. (b) Critical current as a function of $F_\eta$.
}
\end{figure}

\begin{figure}[tb]
\centerline{\includegraphics[width=1\linewidth]{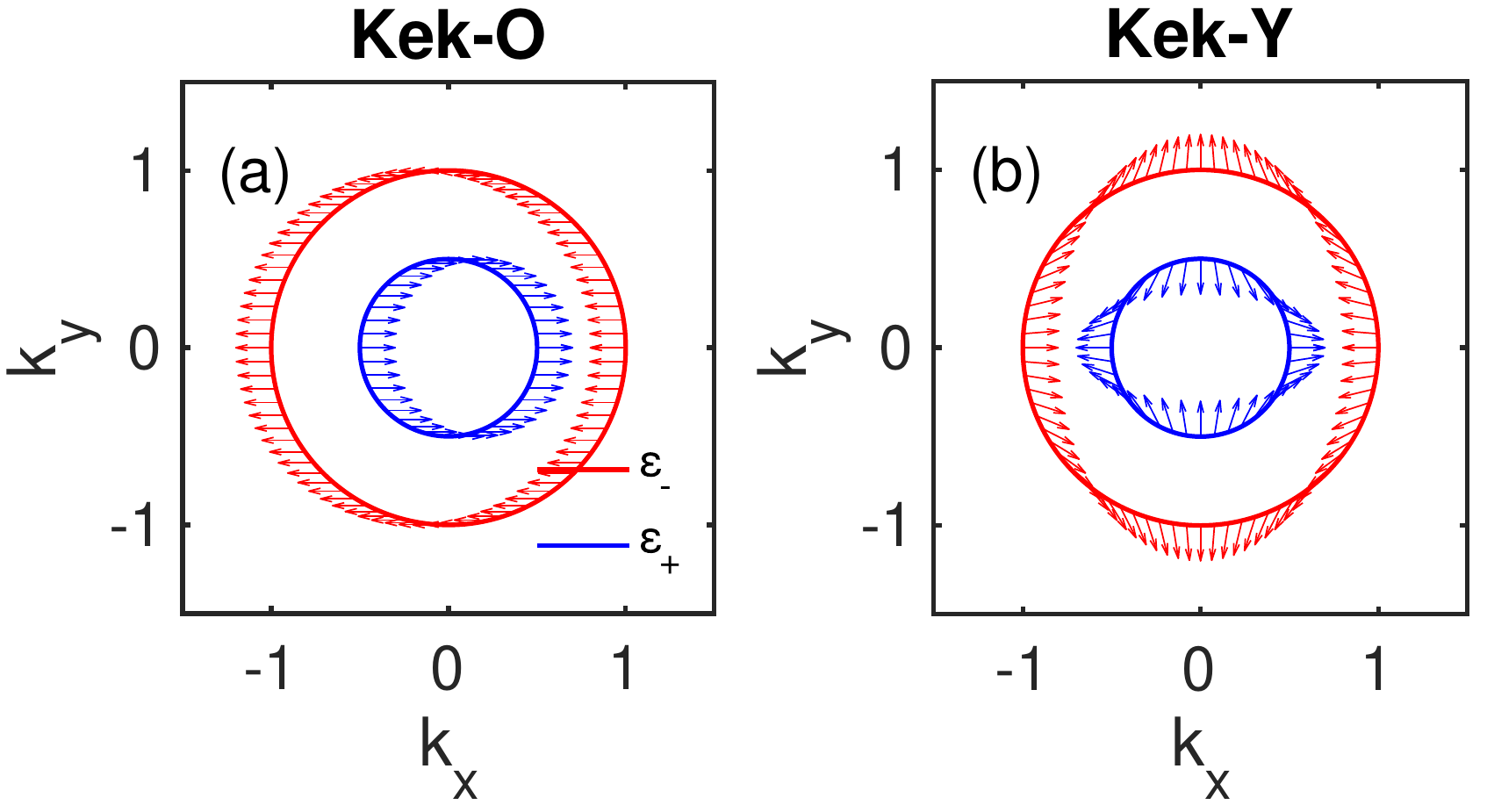}}
\caption{\label{texture}
Schematic diagram of the valley isospin texture for the irradiated Kek graphene at the Fermi surface. Two concentric circles denoted by blue and red lines are the equienergy surfaces for $\varepsilon_\pm$, respectively. The arrows denote the orientation of the valley polarization. The wave vector $k_{x(y)}$ is normalized by $k_F$.
}
\end{figure}

\subsection{Kek-O patterned junction}\label{keko}

With the help of Eq.\ (\ref{josephson}), the current phase relation of the Kek-O Josephson junctions is shown in Figs.\ \ref{f1}(a) and \ref{f1}(b) with different junction lengths $L$ and different illumination parameters $F_\eta$, respectively. One can easily find that a $0$-$\pi$ transition can occur with either increasing $L$ or increasing $F_\eta$.

The signals of the $0$-$\pi$ transitions can also be found in the $L$ dependence of the critical current as shown in Fig.\ (\ref{f22})(a). When there is no light field applied, the critical current exhibits a smooth oscillation behavior as the junction length increasing. The smooth oscillation is attributed to the normal multi-reﬂection at the normal metal/superconductor interfaces and indicates that the junction is always in the $0$-state\cite{PhysRevB.99.045426}. When the light field is on, the critical current curve shows some cuspidal dips, which indicate the $0$-$\pi$ transition. The $0$-$\pi$ transition can also be implied by the dips in the $F_\eta$ dependence of the critical current as shown in Fig.\ \ref{f22}(b).

The $0$-$\pi$ transition in the Kek-O Josephson junction can be understood by a quasiclassical approach employing the Bohr-Sommerfeld quantization condition\cite{Kashiwaya_2000}. By solving the eigenequation of $\mathcal{H}$ in Eq.\ (\ref{total}), one finds that the degenerate conduction band is split into two parts,
\begin{align}
\varepsilon^O_\pm=\sqrt{k_x^2+k_y^2+(F_\eta\pm\mathcal{C}_O)^2}-\mu,
\end{align}
where $k_{x(y)}$ is the longitudinal (transverse) wave vector. For a fixed $k_y$, the scattering states in band $\varepsilon_\pm$ at the Fermi surface are given by  
\begin{align}
&\varphi_+=e^{\pm ik_+x}\begin{pmatrix}
-i e^{i\gamma_+} &\zeta_+^{-1}  & -ie^{i\gamma_+}  & \zeta_+^{-1}
\end{pmatrix}^T\label{wf1},\\
&\varphi_-=e^{\pm ik_-x}\begin{pmatrix}
i e^{i\gamma_-} &-\zeta_-^{-1}  & -ie^{i\gamma_-}  & \zeta_-^{-1}
\end{pmatrix}^T\label{wf2},
\end{align}
with $\zeta_\pm=\sqrt{\frac{\mu+F_\eta\pm\mathcal{C}_O}{\mu-F_\eta\mp\mathcal{C}_O}}$. The longitudinal wave vectors $k_\pm$ are expressed as $k_\pm=\mathcal{P}_\pm\sin\left(\gamma_\pm\right)$ with $\gamma_\pm=\arccos(k_y/\mathcal{P}_\pm)$ and $\mathcal{P}_\pm=\sqrt{\mu^2-(F_\eta\pm\mathcal{C}_O)^2}$.

The matrix structure of the wave functions $\varphi_\pm$ can be rewritten in the tensor product form
\begin{align}
\varphi_\pm\propto\frac{\varphi_K\pm\varphi_{K'}}{\sqrt{2}}\otimes s_\sigma^\pm,
\end{align}
where $\varphi_K=(1,0)^T$ and $\varphi_{K'}=(0,1)^T$ are the the eignvectors of $\tau_z$ and $s_\sigma^\pm=(\mp i e^{i\gamma_\pm},\zeta_\pm^{-1})^T$ represents the sublattice pseudospin component of the wave functions, respectively. The valley isospin texture at the Fermi surface is shown in Fig.\ \ref{texture}(a). It is shown that the $\varepsilon_+$ band is fully valley polarized along $+x$ direction and the $\varepsilon_-$ band is fully valley polarized along $-x$ direction. This is because the Kek-O bond texture couples the valley isospin and the sublattice pseudospin. A light field directly modifying the sublattice pseudospin texture can subsequently change the valley isospin texture as well.

The Josephson current is determined by the Andreev bound states confined in the normal region, which arise from the closed trajectories of the incident electrons and the Andreev reflected holes at the Fermi surface. The $\varepsilon_+$ and $\varepsilon_-$ bands are fully isospin-polarized with opposite orientations, as shown in Fig.\ \ref{texture}(a). In order to form a valley-singlet Cooper pair\cite{PhysRevB.89.064501} in the superconducting regions, an incident electron from the $\varepsilon_+$ band can only be interband reflected to a hole in the $\varepsilon_-$ band so that the intraband Andreev process is totally blockaded. The phase shift can be estimated by the quasiclassical Bohr-Sommerfeld quantization condition, requiring
\begin{align}
\oint kdx+\Omega=l\times2\pi, l\in\mathbb{Z},
\end{align}
where $\Omega=\phi-2\arccos(\varepsilon/\Delta_0)$ is the total Andreev reflection phase acquired at the normal metal/superconductor interfaces. The extra phase accumulated between two interfaces is given by
\begin{align}
&\oint kdx=\pm|k_--k_+|L\label{phase}\notag\\
&\approx \pm\left|\sqrt{\mu^2-(F_\eta+\mathcal{C}_O)^2}-\sqrt{\mu^2-(F_\eta-\mathcal{C}_O)^2}\right|L\notag\\
&=\pm\phi_k,
\end{align}
where the contour integration is along the closed path formed by the classical trajectory of the Andreev reflected particles. The extra phase $\phi_k$ is nonzero in the coexistence of $F_\eta$ and $\mathcal{C}_O$. The total Josephson current can be estimated as
\begin{align}
I=I_+\sin(\phi+\phi_k)+I_-\sin(\phi-\phi_k).\label{pj}
\end{align}
With the assumption that $I_+\approx I_-$, Eq.\ (\ref{pj}) becomes $I\propto\cos\phi_k\sin\phi$, indicating that the Josephson current can be reversed by the extra phase $\phi_k$ with $\pi/2<\phi_k<3\pi/2$. Thus, the $0$-$\pi$ transition can be realized by tuning the junction length or the the illumination parameter, as shown in Fig.\ \ref{f1}.

From Eq.\ (\ref{pj}), one can also find that the critical current $I_c$ is proportional to $\cos\phi_k=\cos(k_+-k_-)L$, which is a periodic function of $L$ with the oscillation period $T_L=2\pi/|k_+-k_-|$. Choosing the same parameters as those in Fig.\ \ref{f22} ($\mu=42$, $\mathcal{C}_O=30$, and $F_\eta=10$), the oscillation period can be estimated as 
\begin{align}
T_L&=\left|\frac{2\pi}{\sqrt{\mu^2-(F_\eta+\mathcal{C}_O)^2}-\sqrt{\mu^2-(F_\eta-\mathcal{C}_O)^2}}\right|\\&\approx0.26,\notag
\end{align}
which is in coincidence with the red solid line in Fig.\ \ref{f22}(a).

\begin{figure}[tb]
\centerline{\includegraphics[width=0.75\linewidth]{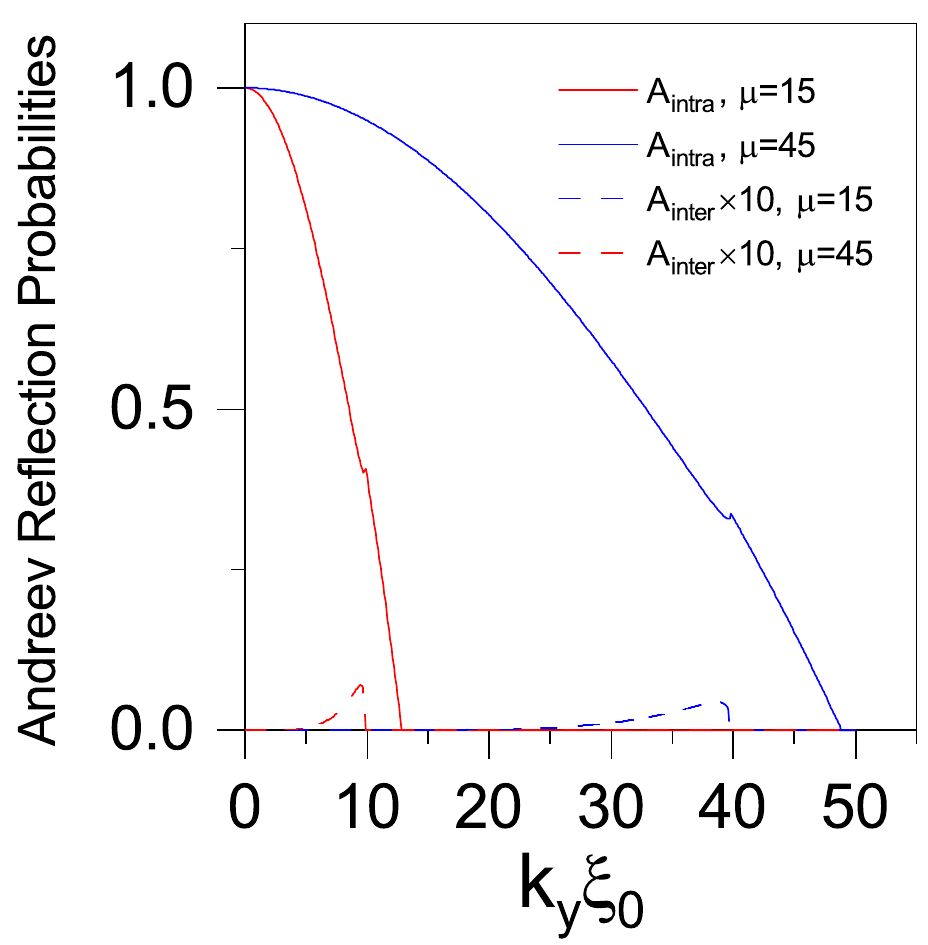}}
\caption{\label{f6}
Andreev reflection probabilities versus the transverse wave vector $k_y$ for Kek-Y junctions. The solid lines indicate the intraband Andreev reflection probabilities. The dashed lines indicate the interband Andreev reflection probabilities.
}
\end{figure}

\begin{figure}[tb]
\centerline{\includegraphics[width=1\linewidth]{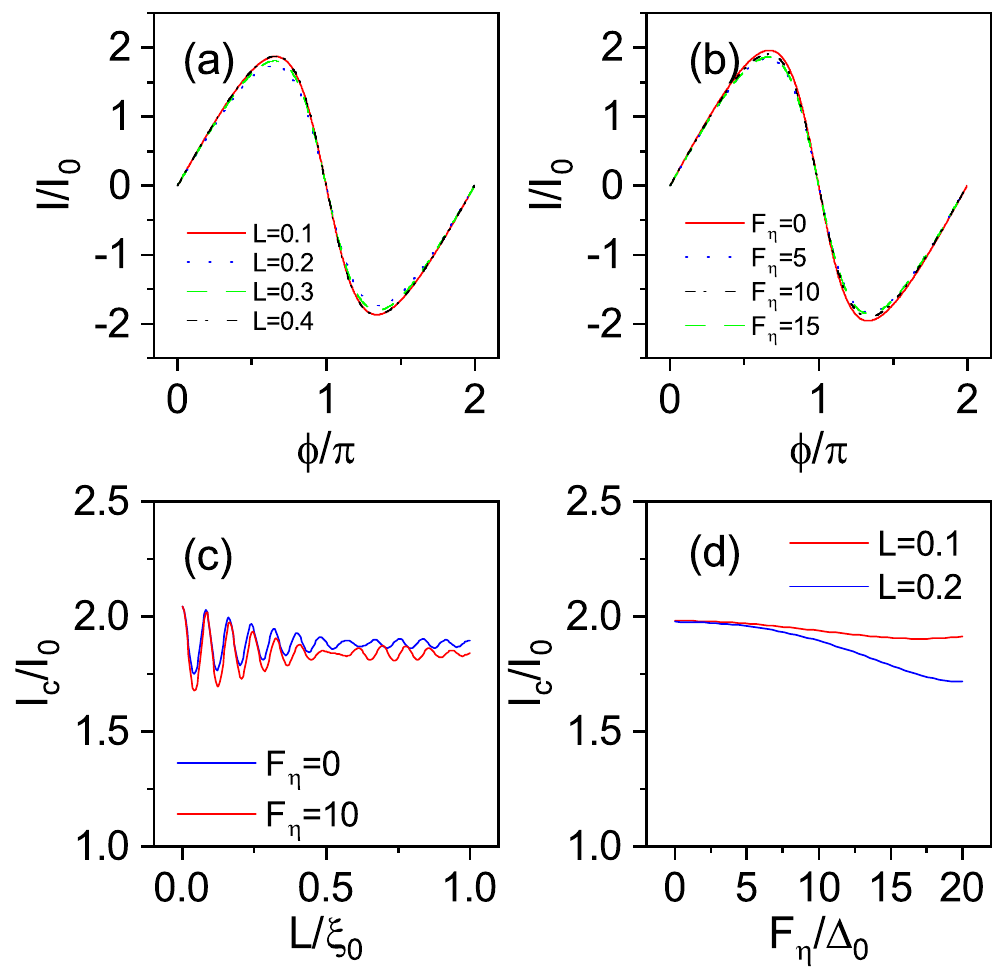}}
\caption{\label{f2}
Current phase relation and critical current in Kek-Y junctions with the parameters $\mu=42$, $\mathcal{C}_Y=0.1$, $T/T_c=0.1$. (a) Current phase relation with $F_\eta=10$. (b) Current phase relation with $L=0.1$. (c) Critical current as a function of $L$. (d) Critical current as a function of $F_\eta$.
}
\end{figure}

\subsection{Kek-Y patterned junction}\label{keky}

\subsubsection{Small-$\mathcal{C}_Y$ regime}
In the small $\mathcal{C}_Y$ regime, i.e., $\mathcal{C}_Y\lesssim0.1$, the term of $\mathcal{O}(\mathcal{C}_Y^2)$ in the Floquet Hamiltonian (\ref{flo}) is negligible. The energy dispersion is given by
\begin{align}
\varepsilon_\pm^Y=\sqrt{F_\eta^2+ k^2}\pm\mathcal{C}_Y k-\mu.
\end{align}
For simplicity, we consider the normal incidence first. At the Fermi surface, the scattering modes for the $\varepsilon_\pm^Y$ bands are given by 
\begin{align}
&\chi_+^\pm=e^{\pm ik'_+x}\begin{pmatrix}
\frac{\mathcal{C}_YF_\eta+\mathcal{A}}{F_\eta-\mu} &\pm1  &\pm\frac{\mathcal{C}_YF_\eta+\mathcal{A}}{F_\eta-\mu} & 1
\end{pmatrix}^T,\label{cx1}\\
&\chi_-^\pm=e^{\pm ik'_-x}\begin{pmatrix}
\frac{-\mathcal{C}_YF_\eta+\mathcal{A}}{F_\eta-\mu} &\mp1  &\mp\frac{-\mathcal{C}_YF_\eta+\mathcal{A}}{F_\eta-\mu}  & 1
\end{pmatrix}^T,\label{cx2}
\end{align}
where $k'_\pm=(\mathcal{C}_Y\mu\mp\mathcal{A})/(1-\mathcal{C}^2_Y)$, $\mathcal{A}=\sqrt{\mu^2-(1-\mathcal{C}^2_Y)F_\eta^2}$ and the subscript $\pm$ of the wave function $\chi$ denotes the $\varepsilon_\pm^Y$ band and the superscript $\pm$ denotes the right and left propagating directions, respectively.

From Eqs.\ (\ref{cx1}) and (\ref{cx2}), one finds that the valley polarization of the right- and left-moving electrons in the $\varepsilon_+$ band is $\langle\chi_+^\pm|\bm\tau\sigma_0|\chi_+^\pm\rangle/\langle\chi_+^\pm|\chi_+^\pm\rangle=(\pm1,0,0)^T$, respectively. Similarly, the valley polarization for the $\varepsilon_-$ band is $(\mp1,0,0)^T$. This valley polarization mismatch in band $\varepsilon_\pm$ prohibits the interband Andreev reflection. Since there is no Fermi momentum splitting as well as the extra phase in the intraband Andreev reflection, the Kek-Y junction always exhibits the 0-state.

For the oblique incidence, the valley isospin is pinned to the $k_x$-$k_y$ plane due to  the valley-momentum locking. The isospin texture at the Fermi surface is plotted in Fig\ \ref{texture}.(b). With a finite $k_y$, the valley isospins in band $\varepsilon_\pm$ are not exactly opposite, leading to a finite interband Andreev reflection. However the main contribution of the Andreev reflection comes from the normal incidence. The interband Andreev reflection activated by the oblique incidence is actually negligible. The interband and the intraband Andreev reflection probabilities are plotted in Fig.\ \ref{f6} for different transverse wave vectors. It is shown that the probability of the interband Andreev reflection is much less than one tenth of the probability of the intraband one. Consequently, the Andreev bound state is dominated by the intraband Andreev reflection without any Fermi-momentum splitting and no 0-$\pi$ transition can be expected. In fact, the Josephson current and the critical current in the Kek-Y junctions can be numerically calculated from Eq.\ (\ref{josephson}) and are plotted in Fig.\ \ref{f2} as functions of the phase difference, the junction length and the illumination parameter, respectively. 
From Figs.\ \ref{f2}(a) and \ref{f2}(b), one always finds the 0-state current phase relation whether or not there is a light field applying. The critical current in Figs.\ \ref{f2}(c) and \ref{f2}(d) always varies smoothly as a function of $L$ and $F_\eta$, still indicating the absence of the 0-$\pi$ transition.

\begin{figure}[tb]
\centerline{\includegraphics[width=1\linewidth]{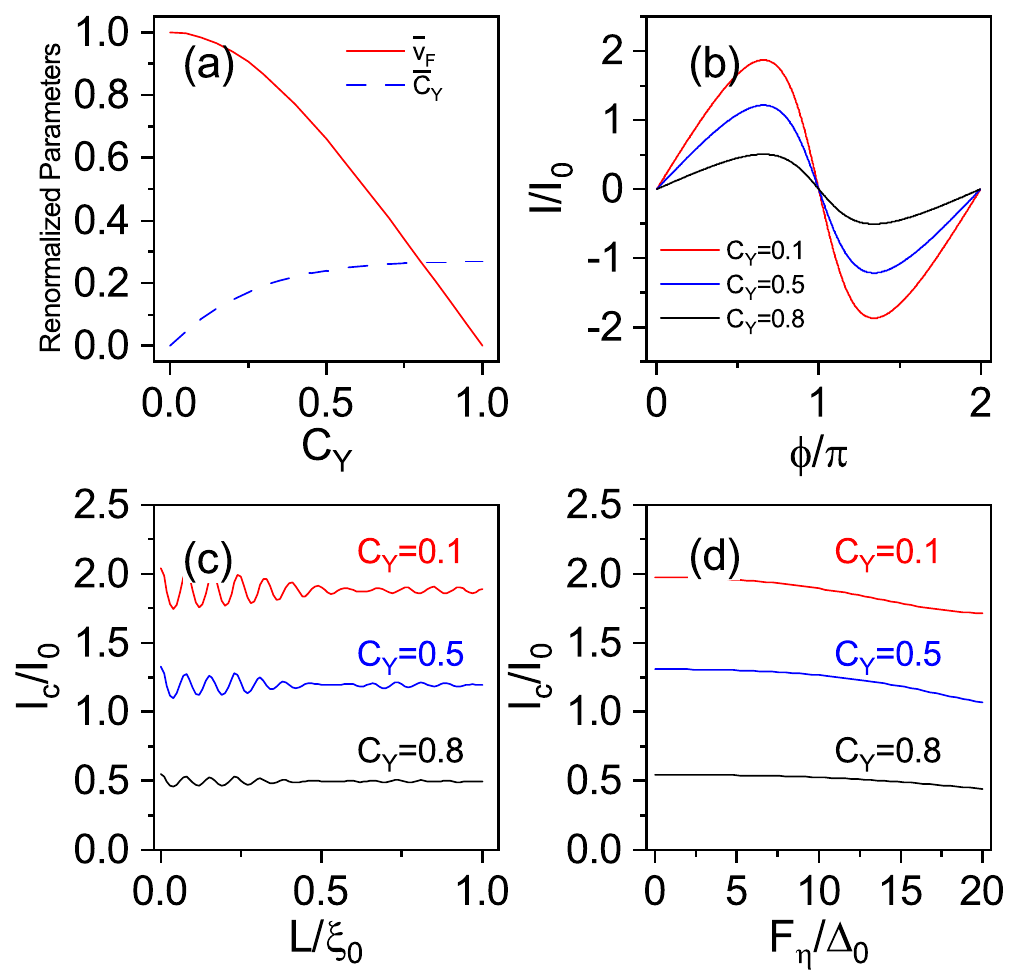}}
\caption{\label{fxt}
(a) Renormalized velocities and renormalized coupling ampiltude as a function of $\mathcal{C}_Y$. (b) Current phase relation with $\mu=42$, $F_\eta=10$, and $L=0.2$. (c) Critical current as a function of $L$ with $\mu=42$, $F_\eta=10$. (d) Critical current as a function of $F_\eta$ with $\mu=42$, $L=0.2$.}
\end{figure}

\subsubsection{Beyond small-$\mathcal{C}_Y$ regime}

Finally, we examine the influence of the Fermi velocity renormalization on the Kek-Y junction when we go beyond the small-$\mathcal{C}_Y$ regime. The low-energy Hamiltonian reads\cite{Gamayun_2018} 
\begin{align}
\overline{\mathcal{H}}_0=\hbar\overline{v}_F\left(\tau_0\bm k\cdot\bm\sigma+\overline{\mathcal{C}}_Y\tilde{\bm k}\cdot\bm\tau\sigma_0\right).\label{rn}
\end{align}
Here the renormalized coupling amplitude $\overline{\mathcal{C}}_Y$ and the Fermi velocity $\overline{v}_F$ are given by 
\begin{align}
&\overline{\mathcal{C}}_Y=\left|\frac{1+2\mathcal{C}_Y-\sqrt{1+2\mathcal{C}_Y^2}}{1+2\mathcal{C}_Y+\sqrt{1+2\mathcal{C}_Y^2}}\right|,\label{A2}\\
\overline{v}_F=v_F&\sqrt{\frac{1+3\mathcal{C}_Y^2+\mathcal{D}(1-3\mathcal{C}_Y^2)+2\mathcal{C}_Y^3(\mathcal{D}-2)}{2\mathcal{D}^2}},\label{A3}
\end{align}
where $\mathcal{D}=\sqrt{1+2\mathcal{C}_Y^2}$.

The renormalized parameters $\overline{v}_F$ and $\overline{\mathcal{C}}_Y$ as a function of $\mathcal{C}_Y$ are shown in Fig.\ \ref{fxt}(a). The renormalized Fermi velocity $\overline{v}_F$ decreases with the increasing of $\mathcal{C}_Y$, which only modifies the normal-state current with the substitution $I_0\rightarrow\overline{I}_0=e\mu\Delta_0 W/\hbar^2\pi \overline{v}_F$. In the regime $0<\mathcal{C}_Y\leq0.1$, one finds that the renormalized coupling amplitude $\overline{\mathcal{C}}_Y$ is approximately the same as $\mathcal{C}_Y$. When $\mathcal{C}_Y>0.1$, $\overline{C}_Y$ increases monotonously and reaches a nearly saturated value. In this regime, the valley isospin is rotated to the $k_z$ axis with a small rotation angle due to the additional term $\overline{\mathcal{C}}_Y^2F_\eta\tau_z\sigma_0$ in the Floquet Hamiltonian. This small rotation of the isospin only modifies the magnitude of the Josephson current but can not change its phase. As a result, the current phase relation and the critical current in the large $\mathcal{C}_Y$ regime is qualitatively the same as those in the small $\mathcal{C}_Y$ regime, as shown in Fig.\ \ref{fxt}.

\section{Conclusions}\label{conclusions}
To conclude, we investigate the dc Josephson effect in Kek patterned graphene. Both the Kek-O and Kek-Y bond texture are considered. It is shown that a light field applied in the normal region provokes entirely different valley isospin texture for the two patterns. The light field and the Kek-O texture together produce a fully valley polarization, leading to the possible $\pi$-state. However, in the Kek-Y graphene, the valley polarization induced by the light field is $k$-dependent and the Josephson junction always exhibits the 0-state.

\section*{Acknowledgements}
This work is supported by the National Key R\&D Program of China (Grant No. 2017YFA0303203) and by the NSFC (Grant No. 11474149).


%

\end{document}